\documentclass[prx,aps,superscriptaddress,footinbib,twocolumn]{revtex4-2}

\usepackage{mathrsfs}
\usepackage{amsfonts}
\usepackage{amssymb}
\usepackage{amsmath}
\usepackage{amsthm}
\usepackage{graphicx}
\usepackage[usenames,dvipsnames]{color}
\usepackage[colorlinks=true,citecolor=blue,linkcolor=magenta]{hyperref}
\usepackage{lmodern}
\usepackage{dsfont}
\usepackage{natbib}
\usepackage{bm}
\usepackage{tabularx}
\usepackage{multirow}
\usepackage{tabularray}
\usepackage{comment}

\usepackage{algcompatible}
\usepackage{algorithm}

\algrenewcommand\algorithmicindent{2em}

\theoremstyle{definition}

\theoremstyle{remark}

\renewcommand{\Re}{\mathrm{Re}}

\newcommand{\abs}[1]{\vert #1 \vert}

\newcommand{\ket}[1]{\vert{ #1 }\rangle}
\newcommand{\bra}[1]{\langle{ #1 }\vert}

\newcommand{\braket}[2]{\langle #1 \vert #2 \rangle}
\newcommand{\mean}[1]{\langle #1 \rangle}

\usepackage{soul}

\renewcommand{\leq}{\leqslant}

\begin{document}

\title{Quantum-enhanced neural networks for quantum many-body simulations}

\author{Zongkang Zhang}
\email{zongkang.zhang.phys@gmail.com}
\affiliation{Graduate School of China Academy of Engineering Physics, Beijing 100193, China}

\author{Ying Li}
\affiliation{Graduate School of China Academy of Engineering Physics, Beijing 100193, China}

\author{Xiaosi Xu}
\email{xsxu@gscaep.ac.cn}
\affiliation{Graduate School of China Academy of Engineering Physics, Beijing 100193, China}

\begin{abstract}
Neural quantum states (NQS) have gained prominence in variational quantum Monte Carlo methods in approximating ground-state wavefunctions. Despite their success, they face limitations in optimization, scalability, and expressivity in addressing certain problems. In this work, we propose a quantum-neural hybrid framework that combines parameterized quantum circuits with neural networks to model quantum many-body wavefunctions. This approach combines the efficient sampling and optimization capabilities of autoregressive neural networks with the enhanced expressivity provided by quantum circuits. Numerical simulations demonstrate the scalability and accuracy of the hybrid ansatz in spin systems and quantum chemistry problems. Our results reveal that the hybrid method achieves notably lower relative energy compared to standalone NQS. These findings underscore the potential of quantum-neural hybrid methods for tackling challenging problems in quantum many-body simulations.
\end{abstract}

\maketitle

\section{Introduction}
The Schrödinger equation lies at the heart of understanding quantum many-body systems, playing a pivotal role in fields such as condensed matter physics, quantum chemistry, and materials science. However, solving the Schrödinger equation for these systems poses a significant challenge due to the exponential growth of the Hilbert space dimension with the number of particles. Exact solutions are confined to relatively small systems, and over the years, significant efforts have been made to develop computational methods to find approximate solutions for specific types of problems. Among these methods, variational approaches, which optimize wavefunctions over parameterized forms, have proven effective~\cite{sandvik2007variational,roggero2013quantum,mahajan2019symmetry}. The accuracy of such approaches, however, depends critically on the ability to describe the wavefunction accurately~\cite{dian2024variational}.

The emergence of artificial neural networks has introduced powerful tools for addressing high-dimensional and intricate problems. Among these advances is the neural quantum state (NQS)~\cite{carleo2017solving,torlai2018neural,luo2019backflow,pfau2020ab,hermann2020deep,bennewitz2022neural}, a framework in which a neural network serves as a variational ansatz to approximate the wavefunction of a quantum system. Due to their high expressive power, neural networks can represent complex quantum many-body states more effectively than traditional variational approaches, such as tensor network states~\cite{glasser2018neural,chen2018equivalence}, matrix product states~\cite{sharir2022neural,wu2023from} or Jastrow–Slater wavefunctions~\cite{pfau2020ab}. Since its introduction, NQS has achieved notable success in computing the ground-state properties of quantum many-body systems~\cite{viteritti2023transformer,nomura2021dirac,ren2023towards,chen2024empowering,ma2024quantum}.

Despite its promise, the NQS approach is not without limitations. Optimization of neural networks often encounters difficulties, including barren plateaus and local minima~\cite{nielsen2015neural,goodfellow2016deep,lange2024from}. Furthermore, scaling NQS methods to larger systems entails significant computational costs~\cite{hermann2023ab}. Additionally, the classical nature of neural networks imposes constraints on their expressivity, limiting their ability to accurately capture the quantum correlations present in many-body wavefunctions~\cite{cai2018approximating,choo2019two,westerhout2020generalization,attila2020neural,passetti2023can,hermann2023ab}.

In recent years, quantum computing has experienced rapid advancements, presenting new opportunities for addressing certain classically-hard problems. One of the principal potential applications of quantum computing is solving quantum many-body problems~\cite{kitaev1995quantum,seth1996universal,low2017optimal,Lin2020nearoptimalground,RevModPhys.92.015003,alexeev2024}. However, the capabilities of current noisy intermediate-scale quantum (NISQ) devices are restricted to small-scale problems due to hardware limitations and the absence of fault-tolerant error correction~\cite{preskill2018quantum,RevModPhys.94.015004}. Nonetheless, quantum states encoded on such devices inherently capture quantum correlations more efficiently than classical representations, offering potential advantages when integrated with classical approaches.

The interplay between quantum computing and neural networks has demonstrated significant potential in a variety of fields, including classification~\cite{havlicek2019}, image processing~\cite{yao2017quantum}, and optimization~\cite{abbas2024challenges}. Hybrid methods often outperform their classical counterparts in specific tasks, showcasing a unique advantage~\cite{callison2022hybrid}. Moreover, numerous studies have illustrated how classical neural networks can assist quantum computing by optimizing quantum circuits~\cite{zhang2022differentiable,liang2023unleashing,nakaji2024generative}, performing post-processing~\cite{zhang2022variational,zhang2025variational}, mitigating noise~\cite{bennewitz2022neural}, learning quantum data~\cite{czischek2022data,hsin2022provably,moss2024enhancing,lange2024transformer}, or improving sampling techniques~\cite{torlai2020precise,yang2024maximizing}. However, relatively few efforts have focused on the reciprocal relationship—how quantum computing can enhance the capabilities of neural networks in solving challenging quantum many-body problems. Bridging this gap is essential to fully harness the complementary strengths of quantum and classical paradigms in advancing computational techniques for quantum systems.

In this work, we propose a hybrid framework that integrates parameterized quantum circuits (PQCs) with neural networks to model the wavefunction of quantum many-body systems. This approach allows for joint optimization of parameters within both the quantum circuit and the neural network, providing the flexibility to adopt sequential or simultaneous optimization strategies tailored to specific problems. Our results demonstrate that this hybrid ansatz not only accelerates convergence but also achieves a notably lower relative error compared to purely classical neural networks, highlighting the potential of quantum-enhanced neural quantum states for overcoming the limitations of existing methods.

\begin{figure*}[htbp]
\centering
\includegraphics[width=0.9\linewidth]{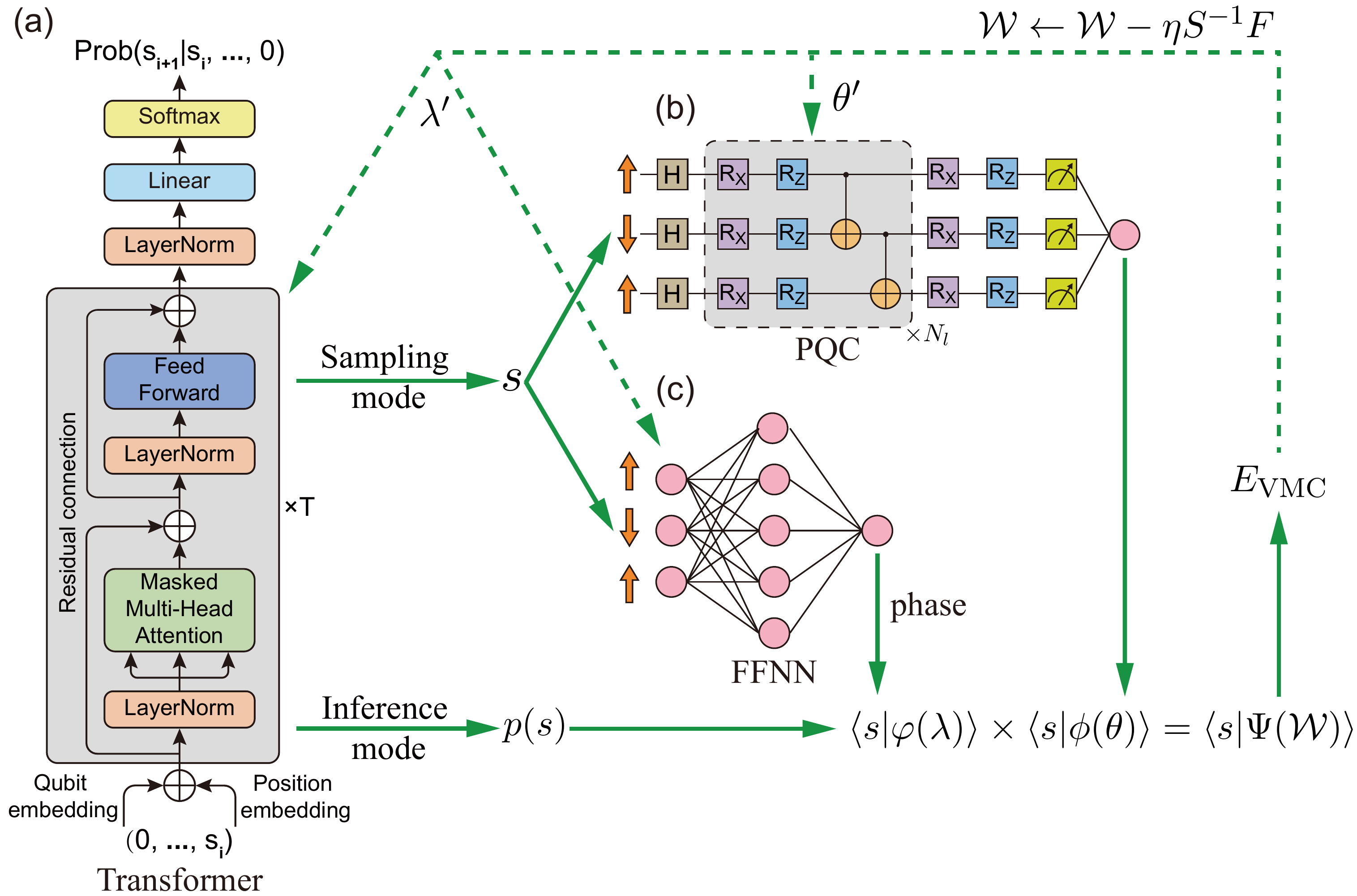}
\caption{Overview and workflow of the quantum-enhanced neural networks. (a) The architecture of the Transformer neural network. (b) The parameterized quantum circuit (PQC), where the rotation gates are defined as $R_X(\theta) = e^{-i\theta X/2}$ and $R_Z(\theta) = e^{-i\theta Z/2}$, with $\theta$ representing the variational parameter to be optimized. (c) The sketch of the feedforward neural network. Our algorithm works as follows: First, the Transformer neural network samples a batch of configurations. Then, the neural networks and the PQCs inference or measure the amplitude and phase for a given configuration. Next, the variational ground state energy, as well as the gradient vector and the quantum fisher matrix, are estimated. Finally, the parameters inside the PQCs and neural networks are optimized based on the stochastic reconfiguration method or the Adam optimizer. This process is repeated until convergence or reaching the maximum number of iterations.}
\label{fig:overview}
\end{figure*}

\section{Quantum-enhanced neural networks}

\subsection{Variational quantum Monte Carlo}
The variational quantum Monte Carlo (VMC) method minimizes the Rayleigh quotient of a given Hamiltonian $H$ by optimizing a parameterized wavefunction $\ket{\Psi(\mathcal{W})}$~\cite{becca2017quantum}. Given access to the complex amplitude $\braket{s}{\Psi(\mathcal{W})}$ for arbitrary configuration $s$, the ground state energy $E_g$ can be estimated as
\begin{align}\label{eq:E_VMC_origin}
E_{\rm VMC} = \frac{\bra{\Psi} H \ket{\Psi}}{\braket{\Psi}{\Psi}} = \Re\mean{E_{loc}(s)},
\end{align}
where $\mean{\bullet} = \mathbb{E}_{s\sim \mathbb{P}(s)} [\bullet]$, $\mathbb{P}(s) = \abs{\braket{s}{\Psi}}^2 / \braket{\Psi}{\Psi}$ and the local energy $E_{loc}(s) = \sum_{s'} \bra{s}H\ket{s'} \braket{s'}{\Psi}/ \braket{s}{\Psi}$. Here, $s$ is sampled according to the probability distribution $\mathbb{P}(s)$. Since the number of $s$ generally increases exponentially with the system size, the denominator may become intractable to sample. For this reason, Markov chain Monte Carlo methods such as the Metropolis-Hastings algorithm is commonly used to obtain a sequence of random samples. In contrast, there is a class of autoregressive neural networks that generates a normarlized wavefunction, enabling direct sampling~\cite{sharir2020deep}. When $s$ is a basis state, the Hamiltonian matrix element $\bra{s}H\ket{s'}$ in the local energy can be computed efficiently~\cite{choo2020fermionic}. Thereby, the computational cost of $E_{loc}(s)$ depends largely on the sparsity of the Hamiltonian. For local Hamiltonians, the number of non-zero elements scales polynomially with the system size. The statistical error of the estimated energy scales as $\sqrt{{\rm Var}[E_{loc}(s)]/B}$, where $B$ is the sample size. While $1/\sqrt{B}$ converges slowly with increasing $B$, VMC has the advantage that, as the wavefunction ansatz converges to an eigenstate, the local energy approaches the corresponding eigenvalue. This results in the variance and the associated statistical error vanishing.

The key point of VMC is to identify a well-designed variational wavefunction. Generally, the greater the expressive power of the variational wavefunction, the more accurate the results. Conventional variational wavefunctions include Hartree-Fock~\cite{hartree1928wave,slater1951}, Jastrow-Slater~\cite{jastrow1955,ceperley1980ground}, configuration interaction~\cite{cramer2013essentials}, coupled cluster~\cite{purvis1982full,RevModPhys.92.015003}, etc~\cite{becca2017quantum}. Recently, NQS with various architectures have been developed~\cite{lange2024from}, including restricted Boltzmann machines~\cite{carleo2017solving}, feedforward neural networks~\cite{cai2018approximating}, convolutional neural networks~\cite{liang2018solving}, recurrent neural networks~\cite{2020recurrent} and Transformer neural networks~\cite{viteritti2023transformer}.

The parameters of the wavefunction can be optimized using various methods, notably stochastic reconfiguration (SR) and Adam~\cite{adam2015}. Both approaches begin with the calculation of gradients $F_i = 2 {\rm Re} [\mean{E_{loc}(s) O_i^*(s)} - \mean{E_{loc}(s)}\mean{O_i^*(s)}]$, where $O_i(s) = \partial \ln{\braket{s}{\Psi}} / \partial \mathcal{W}_i$. SR approximates imaginary time evolution in the parameterized subspace by also evaluating the quantum Fisher matrix $S_{i,j} = {\rm Re} [\mean{O_i^*(s) O_j(s)} - \mean{O_i^*(s)} \mean{O_j(s)}]$~\cite{PhysRevResearch.2.023232}. The parameter update is then performed as
\begin{align}
\mathcal{W} \gets \mathcal{W} - \eta (S + \epsilon I)^{-1} F,
\end{align}
where $\eta$ is the learning rate. Although $S$ is a semi-positive definite covariance matrix, statistical errors can render it non-invertible. A common solution is to add a small positive diagonal matrix $\epsilon I$ to $S$, typically with $\epsilon \sim 10^{-3}$.

Although SR often brings rapid convergence, it also has some challenges. 
The regularization factor $\epsilon$ is sensitive to statistical errors and must be carefully tuned, as overly small values can lead to instability, while larger values reduce optimization accuracy, increasing sampling requirements~\cite{lange2024neural}. In this paper, this trade-off is particularly pronounced as we extend the VMC framework to quantum computing, which involves additional quantum measurement overhead. Furthermore, for systems with a large number of parameters [typically $O(10^3)$], the inverse of $S$ becomes computationally expensive, prompting the development of SR variants~\cite{chen2024empowering,rende2024simple}. Adam, an improved stochastic gradient descent method, avoids these issues by relying solely on gradients without requiring the quantum Fisher matrix $S$. This makes it particularly well-suited for modern large-scale deep network architectures, as it effectively escapes local minima and handles complex optimization landscapes. Combining SR's rapid convergence with Adam's robustness, we use both methods to address their respective limitations in this work.

\subsection{The hybrid quantum-neural wavefunction}
We model the total wavefunction $\ket{\Psi}$ using a combination of the PQC and the neural network. The complex amplitude of a configuration $s$ is defined as 
\begin{align}\label{eq:hybrid_wavefunction}
\braket{s}{\Psi} = \braket{s}{\phi(\theta)} \times \braket{s}{\varphi(\lambda)},
\end{align}
where $\phi(\theta)$ and $\varphi(\lambda)$ represent the quantum and classical wavefunctions, respectively. Here, $s = (s_1, s_2, \cdots, s_{N_q})$ is a binary sequence, where $s_i = 0$ ($1$) represents spin-up (spin-down) in quantum spin systems, or unoccupied (occupied) sites in electronic structure problems. 
This hybrid representation of the wavefunction leverages the expressive power of both the quantum circuit and the neural network, leading to more accurate results. 

For the quantum computing part, we have
\begin{align}\label{eq:QNN}
\ln{\braket{s}{\phi(\theta)}} = f[s;U_1(\theta_1)] +i f[s;U_2(\theta_2)],
\end{align}
where the PQCs $U_1(\theta_1)$ and $U_2(\theta_2)$ are used to generate the amplitude and phase, respectively. Define $f[s;U(\theta)] = \sum_{i=1}^{Nq} c_i \bra{s}U^\dag(\theta) Z_i U(\theta) \ket{s}$, where $N_q$ is the number of qubits.
To estimate $f[s;U(\theta)]$, we apply $U(\theta)$ to the initial state $\ket{s}$ and measure all the qubits in the $Z$ basis. All the expectations $\mean{Z_i}$ can be measured simultaneously, and the process is repeated for sufficient circuit shots $M$. Then $f[s;U(\theta)]$ is the weighted sum of these expectations. Note that except the gate parameters $\theta$, the coefficients $c_i$ are also learnable. In the context of quantum machine learning, such a model is also referred to as a quantum neural network~\cite{abbas2021power}.

The PQC is constructed by alternating layers of single-qubit X and Z rotation gates, along with two-qubit CNOT entanglement gates, commonly referred to as the hardware-efficient ansatz in variational quantum eigensolver~\cite{kandala2017hardware,jerbi2023quantum}:
\begin{align}\label{eq:PQC}
U(\theta) = \prod_{i=1}^{N_l} \left(\prod_{m \neq n} \Lambda_{m,n}\prod_{j=1}^{N_q} R_Z(\theta^Z_{i,j}) R_X(\theta^X_{i,j}) \right) \prod_{k=1}^{N_q} H_k,
\end{align}
where $N_l$ is the number of layers, $H_k$ is a Hadamard gate on the $k$-th qubit and $\Lambda_{m,n}$ is a CNOT gate with the $m$th qubit as the control qubit and the $n$th qubit as the target qubit. 
The entanglement strategies can be customized as needed. For linear entanglement, entanglement gates are applied solely between nearest-neighbor qubits. For full entanglement, each qubit is entangled with every other qubit.
In Fig.~\ref{fig:overview}(b), we present an example of the quantum circuit corresponding to Eq.~(\ref{eq:PQC}) with linear entanglement strategy. In this work, we take $U_1$ and $U_2$ to have the same structure.

The classical neural network part is expressed as:
\begin{align}\label{eq:NQS}
\braket{s}{\varphi(\lambda)} = \sqrt{p(s;\lambda_1)} e^{i \gamma(s;\lambda_2)},
\end{align}
where the probability $p(s;\lambda_1)$ can be modeled by an autoregressive neural network, and the phase $\gamma(s;\lambda_2)$ is represented by another neural network. Commonly used autoregressive neural networks include recurrent neural networks~\cite{2020recurrent}, PixelCNN~\cite{sharir2020deep}, and Transformer neural networks~\cite{bennewitz2022neural,viteritti2023transformer}. In this work, $p(s;\lambda_1)$ is the probability distribution generated by a Transformer neural network, while 
$\gamma(s;\lambda_2)$ is obtained separately from a feedforward neural network. The Transformer neural network has achieved tremendous success in various machine learning domains, including natural language processing, computer vision, and speech processing. 
The Transformer can effectively capture the complex correlations between qubits. More importantly, its autoregressive nature allows for direct sampling of system configurations. Its key advantage lies in the ability to generate all samples in parallel, leading to faster sampling. Additionally, the samples are independent and identically distributed and accurate, resulting in more precise estimates of energy and gradients~\cite{sharir2020deep}. In contrast, non-autoregressive neural networks require sequentially generating Markov chains using algorithms such as Metropolis-Hastings, and discarding a large number of samples through burn-in and thinning. This process is time-consuming and results in less accurate samples.

The structure of the Transformer used in this work is shown in Fig.~\ref{fig:overview}(a).
To predict the first qubit, a leading 
0 is added to the configuration, resulting in the extended configuration $(0, \cdots, s_i)$, where a virtual qubit $s_0 = 0$ is prefixed. The qubit embedding maps the 2-dimensional representation space of qubit states, namely $\ket{0}$ and $\ket{1}$, into a $d$-dimensional embedding space. In large language models, this encoding space is analogous to vocabulary tokens. The positional embedding maps the position of qubits, from $0$ (the prefixed qubit) to $N_q$, into a $d$-dimensional embedding space. Both qubit and position embeddings are learnable and summed together.
After layer normalization, the data proceeds through $h$ heads in parallel. Each head maps the data to a query, key, and value (represented as a three-pronged arrow), with each having a size of $d/h$. The dot product of the query and key represents the attention weights between different qubits. The Transformer enforces its autoregressive nature by preventing $s_i$ from communicating with $s_{i+1}, \cdots, s_{N_q}$, accomplished by masking (setting to zero) the weights connecting them. After performing the weighted aggregation of the values, the outputs of all heads are concatenated back into $d$ dimension. The final output of the masked multi-head attention is linearly projected and added to the original embedding data via a residual connection. After another layer normalization step, the information flows through a fully connected feed-forward network (FFN) with a rectified linear unit (ReLU) as the activation function. Another residual connection is applied. 
The masked multi-head attention and FFN together form a single block, repeated $T$ times. 
After processing through these blocks, the output undergoes a final layer normalization and a linear transformation, mapping the $d$-dimensional embedding data back to the 2-dimensional qubit representation space. A softmax normalization is then applied, producing the normalized conditional probability $p(s_{i+1}|s_i, \cdots, 0)$ due to the autoregressive property. Consequently, $\sum_s p(s) = 1$~\cite{sharir2020deep}, enabling direct sampling. Based on the structure, the total number of parameters in the Transformer can be calculated from $N_q$, $d$ and $T$; further details are provided in Appendix~\ref{app:algorithm}.

The Transformer has two different modes: sampling mode and inference mode~\cite{barrett2022autoregressive}.
In sampling mode, the Transformer generates configurations in parallel according to the probability distribution $p(s)$. As demonstrated in Subsection~\ref{sec:IS}, it serves as a global sampler within our algorithm. In this mode, the Transformer iteratively generates $s_{i+1}$ based on $p(s_{i+1}|s_i, \cdots, 0)$, progressing sequentially from $i=0$ to $N_q-1$ to construct a complete sample. In inference mode, the sampling probability for a given configuration is computed using the Transformer. Specifically, the Transformer computes $p(s_{i+1}|s_i, \cdots, 0)$ for the known sequence $0, \cdots, s_i, s_{i+1}$ from $i=0$ to $N_q-1$. Then, $p(s;\lambda) = \prod_{i=0}^{N_q-1} p(s_{i+1}|s_i, \cdots, 0)$. Notably, for a given configuration, all values of $p(s_{i+1}|s_i, \cdots, 0)$ can be computed simultaneously. In our algorithm, the sampling mode is used exclusively for generating samples, while the inference mode, with its automatic differentiation capabilities, is applied to compute $E_{loc}(s)$ and a portion of $O_i(s)$.

In addition, a feed-forward neural network is used to approximate the phase of the quantum state. This neural network with one hidden layer can be expressed as
\begin{align}\label{eq:FFNN}
\gamma(s;\lambda) = W_2 \cdot  {\rm ReLU}(W_1 \cdot s + b_1),
\end{align}
where ${\rm ReLU}$ is the activation function and $\lambda$ represents all the weights and biases $\{W_1, b_1, W_2\}$. Fig.~\ref{fig:overview}(c) is an illustration of Eq.~(\ref{eq:FFNN}), with the circles and connections denoting the neurons and weights in the neural network, respectively.

\subsection{VMC with Transformer guided importance sampling}\label{sec:IS}
Given the hybrid quantum-neural wavefunction as Eq.~(\ref{eq:hybrid_wavefunction}), the configuration can be sampled according to the probability distribution $\mathbb{P}(s) = \abs{\braket{s}{\phi}}^2 \abs{\braket{s}{\varphi}}^2 / \braket{\Psi}{\Psi}$. However, this approach requires performing Markov chain Monte Carlo sampling, which entails that the quantum computer estimates $\abs{\braket{s}{\phi}}^2$ for configuration $s$ that may ultimately discarded during the sampling process (i.e. burn-in and downsampling). In fact, this problem can be circumvented through importance sampling~\cite{metz2024simulating}. The original sampling probability can be decomposed as
\begin{align}
\mathbb{P}(s) = p(s) \omega(s)
\end{align}
where $p(s) = \abs{\braket{s}{\varphi}}^2 / \sum_s \abs{\braket{s}{\varphi}}^2$ and the importance weight $\omega(s) = \abs{\braket{s}{\phi}}^2 / \mathbb{E}_{s\sim p(s)} [\abs{\braket{s}{\phi}}^2]$. 
Instead of sampling based on $\mathbb{P}(s)$, we can perform importance sampling using $p(s)$, with $\omega(s)$ serving as a modification factor. The Transformer neural network can learn the sampling probability $p(s; \lambda)$, and directly generate a batch of configurations $\{s_{(b)}\}_{b=1}^{B}$, where $B$ is the sample size.
Then, $\omega(s)$ can be estimated on a quantum computer based on the sampled configurations
\begin{align}\label{eq:weight}
\omega(s_{(b)}) = \frac{\abs{\braket{s_{(b)}}{\phi}}^2}{\frac{1}{B} \sum_{b=1}^B \abs{\braket{s_{(b)}}{\phi}}^2 }
\end{align}
Instead of estimating Eq.~(\ref{eq:E_VMC_origin}), the variational energy becomes
\begin{align}
E_{\rm VMC}^H = {\Re} \mean{\omega(s) E_{loc}(s)},
\end{align}
where $\mean{\bullet} = \mathbb{E}_{s\sim p(s)}[\bullet]$ and $E_{loc}(s) = \sum_{s'} \bra{s}H\ket{s'} \braket{s'}{\phi} \braket{s'}{\varphi}/ (\braket{s}{\phi} \braket{s}{\varphi})$. We can estimate $E_{\rm VMC}^H$ by computing the average of $\omega(s) E_{loc}(s)$, i.e.
\begin{align}\label{eq:E_VMC}
\hat{E}_{\rm VMC}^H = \frac{1}{B} {\rm Re} \left[ \sum_{b=1}^B \omega(s_{(b)}) E_{loc}(s_{(b)}) \right].
\end{align}
Similarly for the gradients
\begin{align}\label{eq:F}
\hat{F}_i = & \frac{2}{B} {\rm Re} \bigg[ \sum_{b=1}^B \omega(s_{(b)}) E_{loc}(s_{(b)}) O_i^*(s_{(b)}) \notag\\ &- \sum_{b=1}^B \omega(s_{(b)}) E_{loc}(s_{(b)}) \sum_{b=1}^B\omega(s_{(b)}) O_i^*(s_{(b)}) \bigg].
\end{align}
And the quantum fisher matrix 
\begin{align}\label{eq:S}
\hat{S}_{i,j} = & \frac{1}{B} {\rm Re} \bigg[ \sum_{b=1}^B \omega(s_{(b)}) O_i^*(s_{(b)}) O_j(s_{(b)}) \notag\\ &- \sum_{b=1}^B \omega(s_{(b)}) O_i^*(s_{(b)}) \sum_{b=1}^B\omega(s_{(b)}) O_j(s_{(b)}) \bigg].
\end{align}
To fully specify the computational details of Eqs.~(\ref{eq:F}) and (\ref{eq:S}), we must define the explicit form of $O_i(s)$. The parameters of the hybrid wavefunction present in both the quantum circuit and the neural network, with the complete parameter set denoted as $\mathcal{W} = \{ \lambda, \theta, c \}$. Notice that 
\begin{align}
O_i(s) &= \partial \ln{\braket{s}{\Psi}} / \partial \mathcal{W}_i \notag \\
&= \partial \ln{\braket{s}{\phi}} / \partial \mathcal{W}_i + \partial \ln{\braket{s}{\varphi}} / \partial \mathcal{W}_i,     
\end{align}
allowing us to address these two components separately. For the neural network component in Eq.~(\ref{eq:NQS}), $O^t_{\lambda_i}(s) = \partial \ln{\sqrt{p(s;\lambda)}} / \partial \lambda_i$ and $O^f_{\lambda_i}(s) = i \partial \gamma(s;\lambda) / \partial \lambda_i$ for the parameters of the Transformer and the feedforward neural network, respectively. 
These gradients can be efficiently computed using the built-in backpropagation functionality of deep learning libraries~\cite{rumelhart1986learning}. In the quantum circuit component of Eq.~(\ref{eq:QNN}), $O_{\theta_i}(s) = \partial f[s; U(\theta)] / \partial \theta_i$ and $O_{c_i}(s) = \partial f[s; U(\theta)] / \partial c_i =  \bra{s} U^\dag(\theta) Z_i U(\theta) \ket{s}$ corresponding to the parameters in the amplitude part of the circuit. For the phase part of the circuit, an additional imaginary unit appears in the gradient expressions. Notably, $O_{c_i}(s)$ can be obtained simultaneously when measuring the total expectation $f[s; U(\theta)]$, whereas $O_{\theta_i}(s)$ requires applying the parameter shift rule~\cite{PhysRevLett.118.150503,PhysRevA.98.032309} $\partial f[s; U(\theta)] / \partial \theta_i = (f[s; U(\theta_i + \pi/2)] - f[s; U(\theta_i - \pi/2)]) / 2$. A pseudocode for our algorithm is given in Appendix~\ref{app:code}.

During the optimization process, we can jointly optimize parameters in both the quantum circuit and the neural network, with indices $i(j)$ of $F_i$ and $S_{i,j}$ covering the entire parameter space. Owing to the algorithm's flexibility, we may also first optimize the neural network, then the quantum circuit, or vice versa. When optimizing the neural network or quantum circuit, $i(j)$ only needs to span the parameter space of the corresponding component.
This sequential optimization reduces the number of parameters in every optimization step, avoiding difficulties in inversion due to the large dimension of $S$. Therefore, the hybrid quantum-neural wavefunction can achieve better performance by integrating the strengths of both quantum computing and machine learning. Quantum computing excels at handling complex interactions, compensating for machine learning’s limitations in scalability. Meanwhile, machine learning enhances quantum computing by optimizing resource utilization, improving computational precision, and increasing robustness to noise.

\subsection{Sampling error control}\label{sec:error_control}
In importance sampling, the weight factor affects the variance of the estimator. If the weight factor varies significantly, fluctuations in the results may increase accordingly. As seen from Eq.~(\ref{eq:E_VMC}), the variance of the energy is
\begin{align}
{\rm Var} (E_{\rm VMC}^H) = \frac{1}{B}{\rm Var}\left[\omega(s) \Re(E_{loc}(s))\right].    
\end{align}
Our algorithm still benefits from the property of vanishing variance in VMC as the wavefunction ansatz Eq.~(\ref{eq:hybrid_wavefunction}) converges to the eigenstate of $H$. However, if the difference between the target distribution $\mathbb{P}(s)$ and auxiliary distribution $p(s)$ is large, the importance weights may become highly unstable, leading to inaccurate estimates.

To reduce the sampling error, we should control the range of the weight factor $\omega(s)$. Here, we propose two methods to address this problem. The first approach involves initializing the coefficients $c_i$ and the quantum circuit parameters $\theta_i$ to a small value close to zero. Under this condition, $f[s; U(\theta)] \approx 0$ for arbitrary $s$ such that $\omega(s)$ is optimized from nearly $1$. We observed that this method performed well in numerical simulations. 

We can also apply the activation function ${\rm tanh}$ to control. Specifically, the amplitude $\abs{\braket{s}{\phi}}$ can be changed from $e^{f[s;U_1(\theta_1)]}$ to $e^{a \, {\rm tanh}(f[s;U_1(\theta_1)])}$, where $a$ is a rescaling factor. Then, the amplitude is in the range $[e^{-a}, e^{a}]$. According to Eq.~(\ref{eq:weight}), even in the extreme condition, $\omega(s)$ is less than $e^{4a}$. We note that in practical scenarios, $\omega(s)$ is much smaller than this loose upper bound. When we apply this method, $O_{\theta_i}(s)$ and $O_{c_i}(s)$ will also change accordingly, in addition to the amplitude. It is straightforward that $O_{\theta_i}(s) = a (1 - {\rm tanh}^2 (f[s; U(\theta)])) \partial f[s; U(\theta)] / \partial \theta_i$ and $O_{c_i}(s) = a (1 - {\rm tanh}^2 (f[s; U(\theta)])) \bra{s} U^\dag(\theta) Z_i U(\theta) \ket{s}$.

\subsection{Sampling space reduction}\label{sec:symmetry}
When a quantum system exhibits certain symmetries, the physically valid Hilbert space can be significantly reduced. By modifying the architecture of the Transformer, we can constrain the sampling space to explore, while preserving its autoregressive property. 

For chemical molecules or lattice models such as the Fermi-Hubbard model, the number of electrons $N_e$ and the total electronic spin $S$ are typically conserved. Let $N_{\uparrow}$ and $N_{\downarrow}$ be the number of spin up and down electrons, respectively. Then we have $N_{\uparrow}=N_e/2+S$ and $N_{\downarrow}=N_e/2-S$, which are also conserved. There are many fermion-to-qubit mappings, e.g. parity and Bravyi–Kitaev mappings, etc. Here we choose the Jordan-Wigner (JW) transformation due to its correspondence between the (un)occupation of spin orbital and qubit ($\ket{0}$) $\ket{1}$. The JW transformation converts the second-quantized form of the electronic Hamiltonian to a qubit Hamiltonian. 
It often maps the fermionic indices $j\uparrow$ and $j\downarrow$ into the qubit indices $2j-1$ and $2j$, respectively, where $j=1,\cdots,N_O$ and $N_O$ is the number of spatial orbitals. 

Based on the conservation of $N_{\uparrow}$ and $N_{\downarrow}$, we can mask the Transformer's output probability to avoid invalid outputs. When $i$ is even, the masked probability that the next qubit is in the $\ket{1}$ state-i.e., the spin orbital $(i/2+1)\uparrow$ is occupied-is
\begin{align}
p'(s_{i+1}=1|s_i, \cdots, 0) &= p \times \Theta( N_{\uparrow} - \sum_{k=1}^{i/2} s_{2k-1} ),
\end{align}
where $p$ is the original probability and $\Theta(x)$ denotes the Heaviside step function, defined as $\Theta(x)=1$ for $x > 0$ and $\Theta(x)=0$ for $x\leq 0$. Conversely, the masked probability that the next qubit is in the $\ket{0}$ state is
\begin{align}
&p'(s_{i+1}=0|s_i, \cdots, 0) \notag\\
&= p \times \Theta\left[ (N_O-N_{\uparrow}) - \sum_{k=1}^{i/2} (1-s_{2k-1}) \right].
\end{align}
Similarly, when $i$ is odd, we have
\begin{align}
p'(s_{i+1}=1|s_i, \cdots, 0) = p \times \Theta( N_{\downarrow} - \sum_{k=1}^{\lfloor i/2 \rfloor} s_{2k} ), 
\end{align}
\begin{align}
&p'(s_{i+1}=0|s_i, \cdots, 0) \notag\\
&= p \times \Theta\left[ (N_O-N_{\downarrow}) - \sum_{k=1}^{\lfloor i/2 \rfloor} (1-s_{2k}) \right].
\end{align}
After this step, the masked probabilities should further be normalized by dividing $p'(s_{i+1}=1)$ and $p'(s_{i+1}=0)$ by the normalization factor $p'(s_{i+1}=1)+p'(s_{i+1}=0)$. 
By considering symmetries, the dimension of the sample space is reduced from $2^{N_e}$ to $\binom{N_O}{N_{\uparrow}}\binom{N_O}{N_{\downarrow}}$. This dimensional reduction improves the efficiency of the sampling and the accuracy of energy estimation.

\section{Results}\label{sec:results}
In this section, we present numerical simulations of our algorithm. First, we evaluate the energy error and statistical error across different Monte Carlo sampling sizes and quantum circuit measurement counts. Second, we analyze the trend of relative error as system size increases. Third, through sequential optimization, we demonstrate enhanced expressive power with the addition of quantum circuit layers. Finally, we show that the quantum-neural hybrid ansatz can outperform the classical neural quantum state alone.

We address the problem of finding the ground state for two classes of physical systems: spin models and chemical molecules. A representative spin model is the anti-ferromagnetic Heisenberg (AFH) model, with its Hamiltonian given by
\begin{align}
H_{\rm AFH} = J \sum_{<i,j>} (\sigma_i^X \sigma_j^X + \sigma_i^Y \sigma_i^Y + \sigma_i^X \sigma_i^X),
\end{align}
where $J$ is the interaction strength and $\sigma_i^X$, $\sigma_i^Y$ and $\sigma_i^Z$ are Pauli matrices at the $i$-th spin. For our numerical simulations, we consider a one-dimensional AFH model with periodic boundary conditions (PBC) and set $J=1$.

The electronic Hamiltonian of chemical molecules, expressed in the second-quantized form under fermionic anti-commutation relations and a specific orbital basis set, is given as~\cite{RevModPhys.92.015003}
\begin{align}\label{eq:H_e}
H_{e} = \sum_{ij} h_{ij} c_i^\dagger c_j + \frac{1}{2} \sum_{ijkl} h_{ijkl} c_i^\dagger c_j^\dagger c_k c_l,
\end{align}
where $c_j^\dagger$ and $c_j$ are the fermionic creation and annihilation operators, respectively. The coefficients $h_{ij}$ and $h_{ijkl}$ represent the one-body and two-body integrals, respectively.
We note that the Hamiltonian of the Fermi-Hubbard model, a cornerstone in the study of strongly correlated systems, is a special case of Eq.~(\ref{eq:H_e}). To map the fermionic operators $c_j^\dagger$ and $c_j$ to Pauli operators, we apply the JW transformation, defined as
\begin{align}
\left\{
\begin{aligned}
c_j^\dagger &= \frac{1}{2} (X_j - i Y_j) \prod_{k<j} Z_k, \\
c_j &= \frac{1}{2} (X_j + i Y_j) \prod_{k<j} Z_k,
\end{aligned}
\right.
\end{align}
where $X_j$, $Y_j$, $Z_j$ are Pauli operators.
This transformation rewrites the second-quantized Hamiltonian in terms of Pauli operators, enabling its compatibility with spin-based neural network state representations~\cite{choo2020fermionic}. In this work, we use OpenFermion~\cite{McClean2020} to generate the electronic Hamiltonian and apply the JW transformation.

\begin{figure}[htbp]
    \centering
    \includegraphics[width=0.95\linewidth]{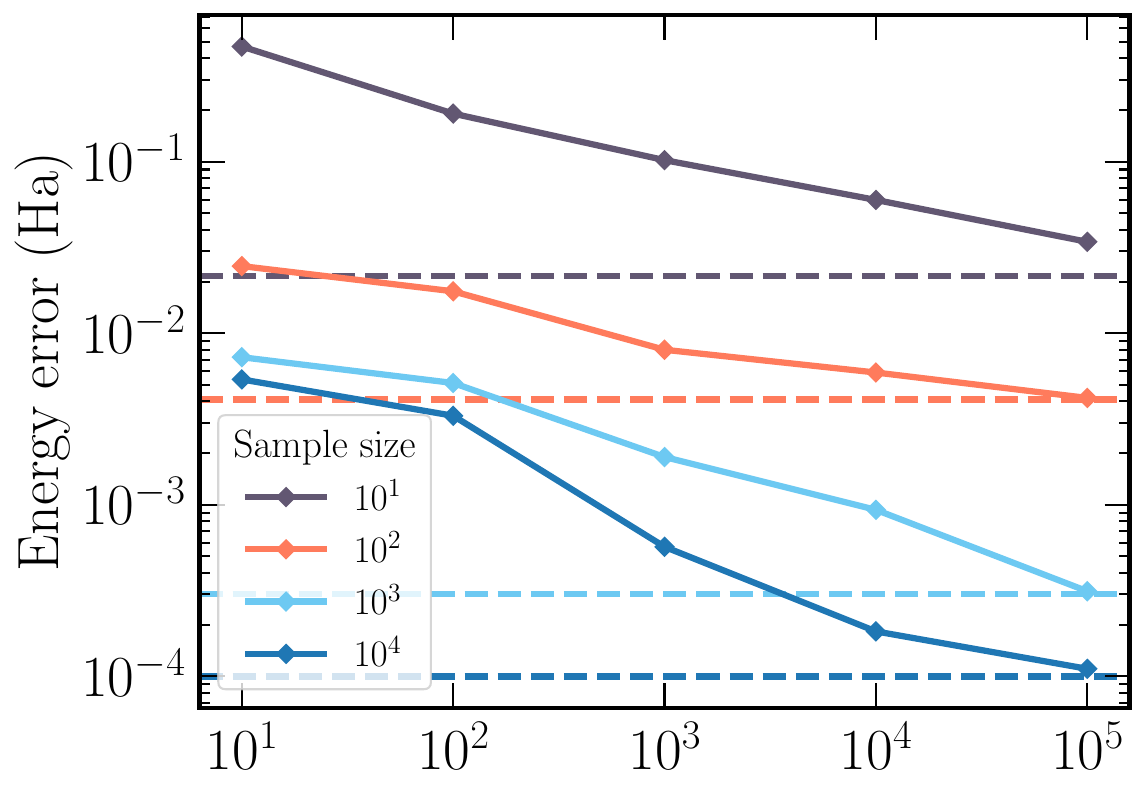}
    \vspace{0.cm} 
    \includegraphics[width=0.95\linewidth]{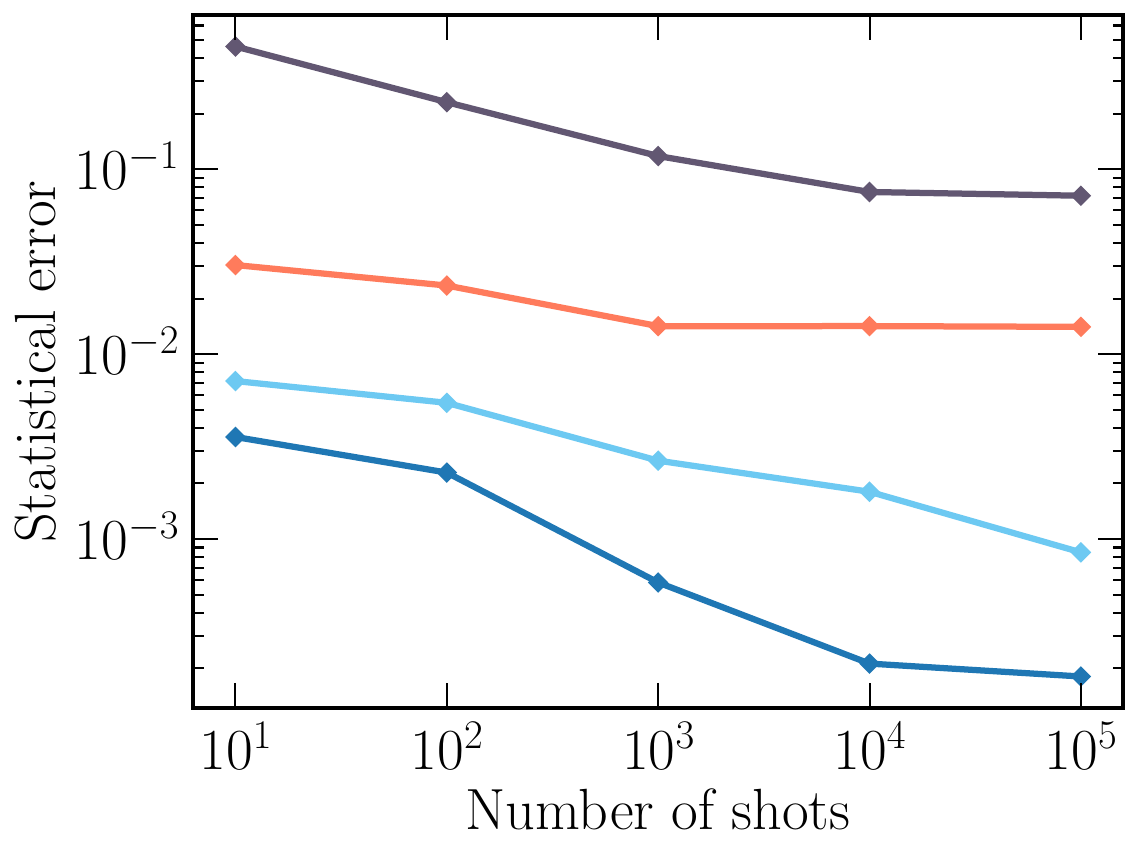}
    \caption{Energy error (top) and statistical error (bottom) of our algorithm for different sample sizes and shot numbers. The model studied is a 4-spin AFH chain with PBC. The SR method is employed for optimization. The energy error represents the average deviation from the exact energy, computed over 100 random seeds, while the statistical error corresponds to the standard deviation of results across the same seeds.  Dashed lines indicate the estimated energy assuming an infinite number of shots, i.e., no shot noise in quantum measurements. The Transformer has an embedding dimension $d=8$, $h=4$ heads and $T=2$ blocks, and the feedforward neural network contains one hidden layer of 16 neurons. The quantum circuit employs 2 layers with the full entanglement strategy.}
    \label{fig:statistical_error}
\end{figure}

In our algorithm, statistical errors arise from both Monte Carlo sampling from the Transformer and the finite number of shots in the quantum circuit. These errors, in turn, impact the accuracy of the final results. Using the two methods described in Sec.~\ref{sec:error_control}, we can systematically control statistical errors to achieve more accurate results. In Fig.~\ref{fig:statistical_error}, we show the energy errors and statistical errors of our algorithm for various sample sizes $B$ and shot numbers $M$. We select sufficiently large hyperparameters to minimize the energy bias caused by the limited expressive power of the wavefunction ansatz. As the sample size and shot number increase, the accuracy of the results improves, while the variance decreases accordingly. The total measurement cost of quantum circuits scales as $\mathcal{O}(BM)$. In our simulations, we observe that the algorithm achieves desirable performance when $B$ and $M$ are within the range of $\mathcal{O}(10^3-10^4)$. When the shot number approaches $\mathcal{O}(10^5)$, the results converge to those obtained from the simulation of the quantum circuit without shot noise.

\begin{figure}[htbp]
\centering
\includegraphics[width=\linewidth]{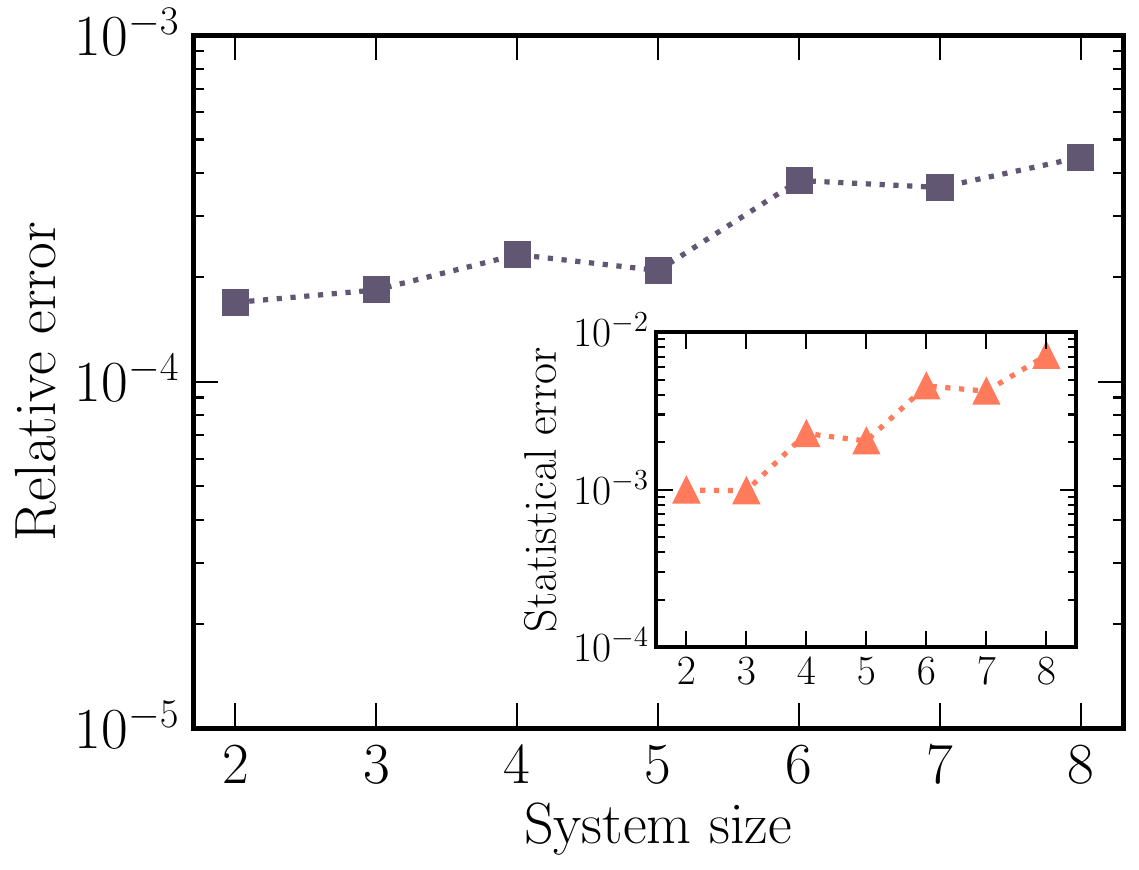}
\caption{Relative energy error $\abs{(E_{\rm VMC}^H-E_g)/E_g}$ and statistical error for the AFH chain with PBC for systems of $2$ to $8$ spins. Both the sample size and shot number are set to $10^3$. The relative errors represent averages over 100 random seeds, while the statistical errors correspond to the standard deviation across the same set of seeds. The Transformer has an embedding dimension $d=8$, $h=4$ heads and $T=2$ blocks, and the feedforward neural network contains a single hidden layer of 16 neurons. The quantum circuit consists of 2 layers employing the linear entanglement strategy. }
\label{fig:relative_error_vs_system_size}
\end{figure}

Next, we evaluate the scalability of the algorithm by computing the ground states of an AFH chain with 2 to 8 spins. Except for the input dimension, which increases with system size, the hyperparameters of the neural network and quantum circuit stay unchanged. Fig.~\ref{fig:relative_error_vs_system_size} shows the relative energy errors and the statistical errors of the results for different system sizes. As the system size increases, the final relative errors are  around $3 \times 10^{-4}$ and show only a slight increase, demonstrating the algorithm's effectiveness in handling larger systems. 
Moreover, we observe that the results for odd $N_q$ are marginally better than those for $N_q-1$. This can be explained by the fact that in the AFH model with PBC, the ground state is fourfold degenerate for odd $N_q$ due to spin and translational symmetries, while it is non-degenerate for even $N_q$. The degeneracy of the ground state for odd $N_q$ simplifies the optimization process, potentially leading to better results.

\begin{figure}[htbp]
\centering
\includegraphics[width=\linewidth]{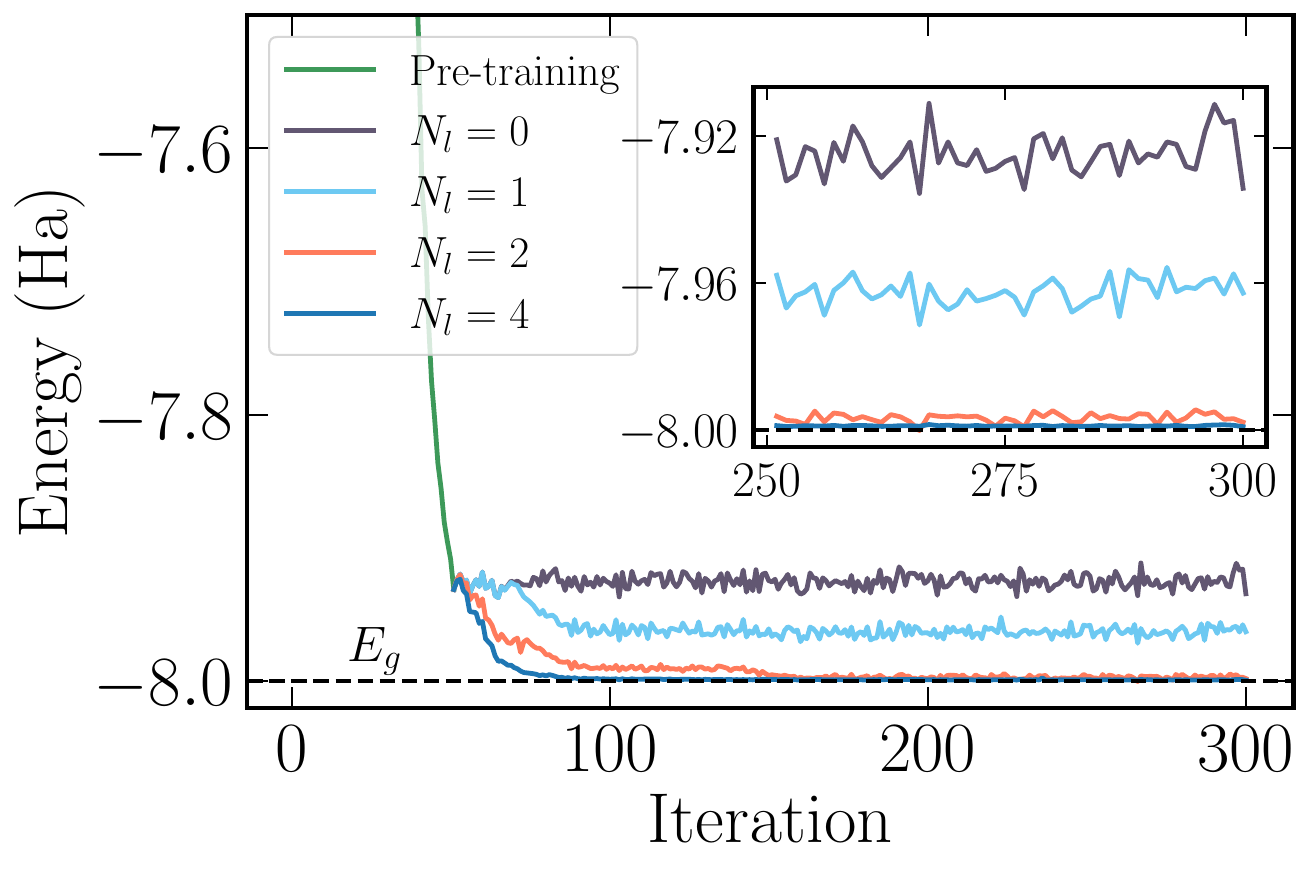}
\caption{Sequential optimization for quantum circuits with varying numbers of layers. The neural network is first pre-trained to provide a good initial approximation of the energy. Subsequently, quantum circuits with different numbers of layers are optimized to refine the results. The sample size is set to $10^4$ and the quantum measurements are assumed to be exact. The Transformer has an embedding dimension $d=4$, $h=2$ heads and $T=1$ block and the feedforward neural network contains a single hidden layer with $16$ neurons. The quantum circuit employs the full entanglement strategy.}
\label{fig:segmented_opt}
\end{figure}

By employing the sequential optimization method, we can fix the neural network parameters and investigate the expressive power of quantum circuits with varying numbers of layers. 
The optimization process can be divided into two stages. 
First, we optimize the NQS described by Eq.~(\ref{eq:NQS}) to obtain a good approximation of the ground state. Then, by fixing the parameters of the neural network, we further refine the wavefunction ansatz using a PQC. Fig.~\ref{fig:segmented_opt} presents the sequential optimization curves for various quantum circuit layers. When $N_l=0$, the quantum circuit contains no entanglement gates and consists solely of a single layer of rotation gates, resulting in almost no improvement. 
As $N$ increases from $1$ to $4$, the relative energy error continuously decreases, reaching approximately $10^{-4}$. This trend demonstrates that increasing the number of layers in the quantum circuit enhances its expressive power. Furthermore, we observe a significant reduction in fluctuations as $N_l$ increases. This improvement is attributed to the wavefunction ansatz more closely approximating the ground state, benefiting from the property of vanishing variance.

Beyond spin models, our algorithm is also applicable to chemical molecules. As an example, we consider the STO-3G basis set for {\rm LiH}, where lithium contributes the $\{1s, 2s, 2p_x, 2p_y, 2p_z\}$ orbitals and the hydrogen contributes the $\{1s\}$ orbital. Using the one-particle reduced density matrix (1-RDM), the occupation numbers of the orbitals can be determined. Orbitals with occupation numbers close to 0 or 1 are classified as virtual or core orbitals, respectively. By applying this method, the number of spin-orbitals is reduced from 12 to 6. Additionally, the masked probability procedure described in Sec.~\ref{sec:symmetry} is employed to ensure that the sampling process rigorously preserves physical symmetries. It is important to note that once specific core orbitals are fixed, the number of electrons $N_e$ is restricted to the remaining active electrons.

\begin{figure}[htbp]
\centering
\includegraphics[width=\linewidth]{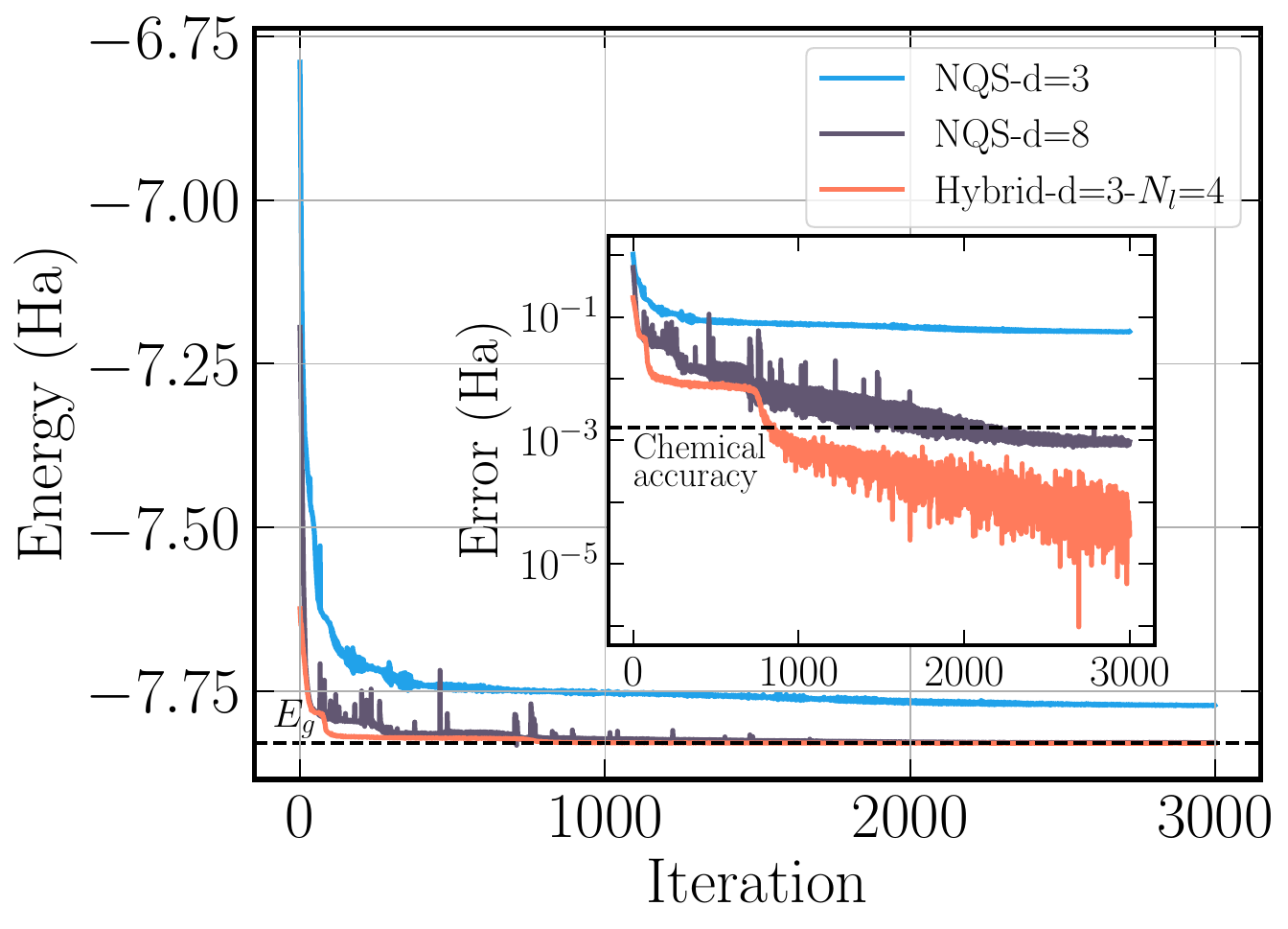}
\caption{Energy convergence curve for the NQS and the hybrid quantum-neural method for a 6-qubit {\rm LiH} molecule with an interatomic distance of $2.4$ \AA. The blue curve represents the performance of the NQS with an embedding dimension of $d=3$, while the orange curve shows the result of our hybrid method, which incorporates a quantum circuit with $N_l=4$. The hyperparameters for these configurations are detailed in Table~\ref{tab:hyperparameters} of Appendix~\ref{app:parameters}. The gray curve corresponds to an NQS with an expanded parameter set, where the Transformer architecture utilizes an embedding dimension of $d=8$, $h=4$ heads and $T=2$ blocks, and the feedforward neural network comprises two hidden layers with 64 neurons each. The inset displays the energy error on a logarithmic scale, with chemical accuracy serving as the benchmark.}
\label{fig:hybrid_vs_nqs-LiH-distance=2.4}
\end{figure}

To highlight the potential of quantum computing in enhancing the expressive power of NQS, we compare our hybrid quantum-neural method with a standalone NQS. Our results demonstrate that when the classical neural network's expressive power is constrained—either by its architecture or a limited number of parameters—a PQC can effectively augment the wavefunction ansatz.
Fig.~\ref{fig:hybrid_vs_nqs-LiH-distance=2.4} presents the energy convergence curves for the {\rm LiH} molecule at an interatomic distance of $2.4$ \AA\ under varying hyperparameter settings. An NQS with $d=3$ employs a Transformer with 179 parameters and a feedforward neural network with 256 parameters, achieving an average energy error of $5.8 \times 10^{-2}$ Ha. However, when combined with a four-layer PQC containing 60 additional parameters, the hybrid method surpasses the chemical accuracy threshold ($1.6\times 10^{-3}$ Ha) after $10^3$ iterations and achieves an average error of $6.0 \times 10^{-5}$ Ha by the end. For comparison, we also optimized an NQS with significantly more parameters, comprising a Transformer with 1802 parameters and a feedforward neural network with 4672 parameters—over ten times the parameter count of the smaller NQS. This larger NQS achieves an energy error of $9.4 \times 10^{-4}$ Ha, which slightly exceeds the chemical accuracy but still falls short of the hybrid method's performance.
These findings demonstrate that the hybrid method converges more rapidly and achieves greater accuracy in variational energy evaluations compared to NQS alone, while requiring significantly fewer parameters. Similar results are observed for the AFH model, as illustrated in Fig.~\ref{fig:hybrid_vs_nqs-Heisenberg-Nq=7} in Appendix~\ref{app:AFH}.

\begin{figure}[htbp]
\centering
\includegraphics[width=\linewidth]{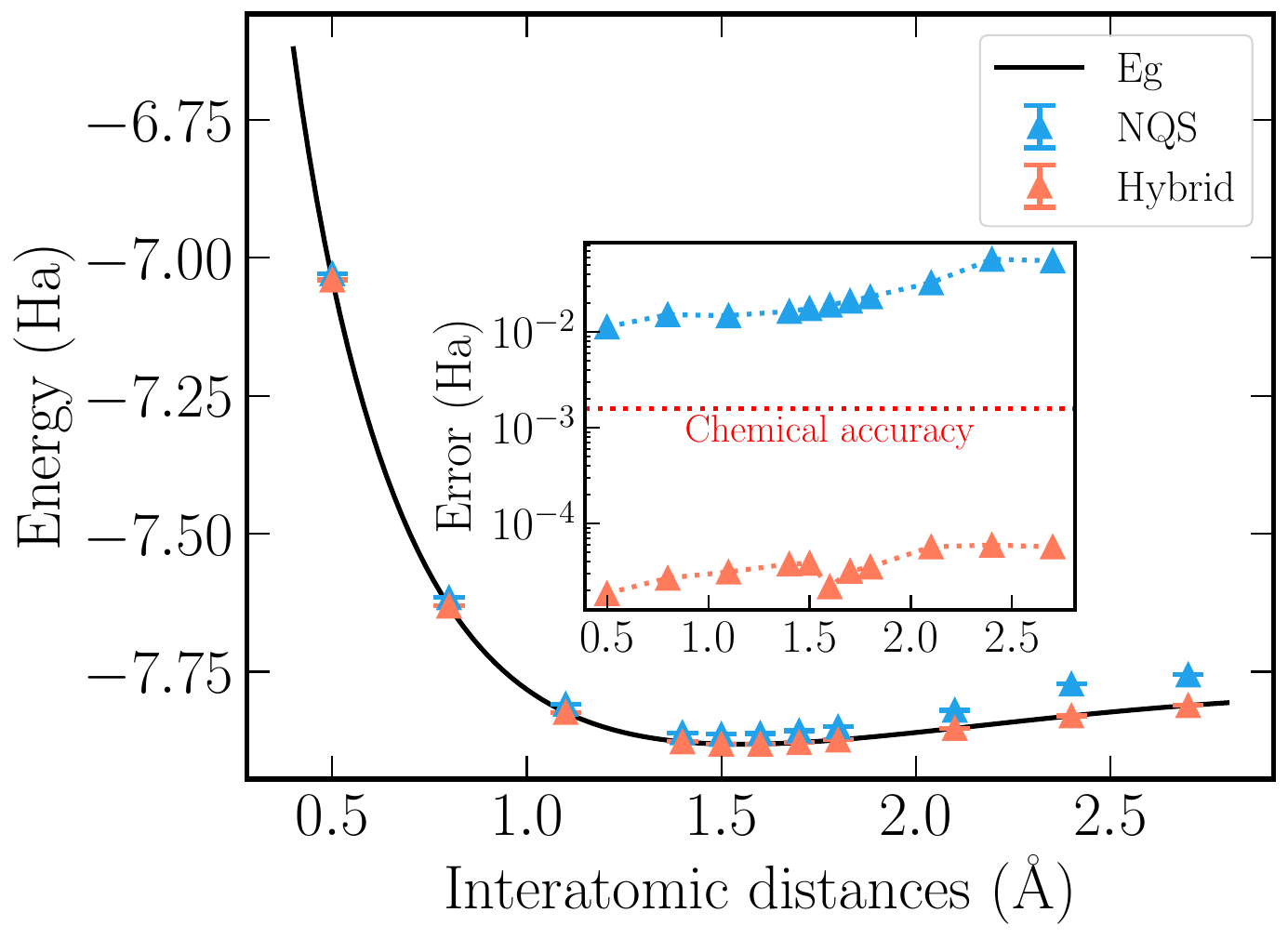}
\caption{Potential energy surface of the {\rm LiH} molecule calculated using the NQS (blue) and our hybrid quantum-neural method (orange). Using the STO-3G basis, the first spatial orbital is designated as the core orbital, while the fourth and fifth orbitals are treated as virtual orbitals, based on the 1-RDM analysis performed with OpenFermion~\cite{McClean2020}. For each interatomic distance, the variational energy is computed as the mean of the final 100 iterations, with error bars indicating the corresponding standard deviation. The inset shows the energy error on a logarithmic scale, with the red dashed line representing chemical accuracy as a benchmark. The hyperparameters are provided in Table~\ref{tab:hyperparameters} of Appendix~\ref{app:parameters}.}
\label{fig:LiH}
\end{figure}

In Fig.~\ref{fig:LiH}, we compare the performance of the NQS and our hybrid method in calculating the potential energy surface of {\rm LiH} with interatomic distances ranging from $0.5$ \AA\ to $2.7$ \AA. 
Due to the relatively small number of parameters in the NQS, the resulting energy error is above $10^{-2}$ Ha. By incorporating a shallow PQC, our hybrid method achieves an energy error reduction of nearly three orders of magnitude, bringing it well below chemical accuracy.
Here, to highlight the enhancement of expressive power provided by quantum computing, we selected neural networks with a constrained number of parameters for this study. In principle, neural networks with larger hyperparameters can achieve greater expressive power. However, for certain strongly correlated models, single neural networks face intrinsic limitations in accuracy~\cite{hermann2023ab,ibarra2024autoregressive}. In such scenarios, quantum computing can serve as a valuable complementary approach.

\section{Discussion and Conclusion}
In this work, we introduce a hybrid quantum-neural method for solving quantum many-body problems in discrete systems, such as spin models and second-quantized chemical molecules. The approach is built on a hybrid quantum-neural wavefunction, combining a neural quantum state (NQS)—incorporating a Transformer neural network and a feedforward neural network—with a parameterized quantum circuit (PQC). By employing an importance sampling technique that uses the Transformer as a direct sampler, the method improves efficiency and addresses the limitations of traditional Markov chain Monte Carlo sampling. To further enhance performance, we use two variance reduction strategies: (1) constraining the wavefunction amplitude using activation functions and (2) leveraging problem-specific symmetries to refine the Transformer's sampling probabilities, effectively reducing the sampling space and improving accuracy.

Numerical simulations demonstrate the effectiveness of the proposed method. Incorporating a PQC enhances the expressive power of the NQS, allowing it to achieve higher accuracy with significantly fewer parameters compared to conventional NQS. Additionally, our analysis of energy and statistical errors shows systematic performance improvements with increased sample size, quantum circuit depth, and the number of circuit shots. The method also exhibits favorable scalability, with errors increasing modestly as the system size grows.

This work opens several avenues for future exploration. While we employ a general hardware-efficient PQC in this study, designing problem-specific PQCs tailored to particular systems could yield larger benefits. Similarly, expanding beyond Transformers and feedforward neural networks to explore more advanced and deeper NQS architectures may extend this approach to larger and more complex physical systems, where the hybrid quantum-neural method could provide greater advantages. Additionally, investigating the noise resilience of this method on quantum hardware will be important for realizing the full potential of this approach in practical applications.

\begin{acknowledgments}
We implement the neural networks in PyTorch~\cite{paszke2019pytorch}, construct the quantum circuits with Qiskit~\cite{qiskit2024}, and generate the molecular Hamiltonians using OpenFermion~\cite{McClean2020}. The source codes for the numerical simulation are available at GitHub~\cite{code}. The authors acknowledge discussions with Ji Chen. This work is supported by Innovation Program for Quantum Science and Technology (Grant No. 2023ZD0300200), National Natural Science Foundation of China (Grants No. 12225507 and No. 12088101),  and NSAF (Grants No. U1930403, No. U2330201 and No. U2330401).
\end{acknowledgments}

\clearpage
\bibliography{reference}

\clearpage
\onecolumngrid
\appendix

\section{Number of parameters in the Transformer neural network}\label{app:algorithm}

The total number of parameters of the Transformer neural network, as shown in Fig.~\ref{fig:overview}(a), can be determined based on the number of qubits $N_q$, the embedding dimension $d$, and the number of blocks $T$. In the embedding layer, the qubit and position embedding has $2d$ and $(N_q + 1)d$ parameters, respectively. 
The layernorm has $2d$ parameters.
In the masked multi-head attention layer, each head has $3\times d \times d/h$ parameters, resulting in a total of $3d^2$ parameters across all heads. Additionally, there are $d^2+d$ parameters for the linear projections.
The feedforward layer consists of two hidden layers, each with $4d$ neurons, resulting in a total of $8d^2+5d$ parameters.
The linear layer has $2d+2$ parameters. Therefore, the total number of parameters is
\begin{align}
& 2d + (N_q+1)d + (2d + 3d^2 + d^2+d + 2d + 8d^2+5d)\times T + 2d + 2d+2 \notag \\
= & 12Td^2+(10T+N_q+7)d+2.
\end{align}

\section{Pseudocode for the hybrid quantum-neural networks}\label{app:code}

\begin{figure}[htbp]
\begin{algorithm}[H]
\caption{VMC with Hybrid Quantum-Neural Wavefunction}
\label{alg:vmc}
\begin{algorithmic}[1]
    \REQUIRE Hamiltonian $H$ and hybrid quantum-neural wavefunction $\braket{s}{\Psi(\theta, \lambda)}$ defined in Eq.~(\ref{eq:hybrid_wavefunction}), sample size $B$, shot number $M$, learning rate $\eta$, and regularization constant $\epsilon$.
    \ENSURE Estimation of ground state energy $E_{\rm VMC}^H$.
    
    \FOR{$i = 1$ to \textit{iterations}} 
        \STATEx \hspace{2em}\COMMENT{Transformer-guided importance sampling}
        \STATE Use the Transformer in sampling mode to generate a batch of samples $\{ s_{(b)} \}_{b=1}^B$ in parallel
        \FOR{$b = 1$ to $B$} 
            \STATEx \hspace{4em}\COMMENT{Record the importance weight and local energy for each sample}
            \STATE Estimate $\{ \omega(s_{(b)}) \}_{b=1}^B$ and $\{ E_{loc}(s_{(b)}) \}_{b=1}^B$
            \STATEx \hspace{4em}\COMMENT{Compute the derivative of the neural network parameter by backpropagation}
            \STATE Estimate $\{ O_{\lambda_i}(s_{(b)}) = \partial \ln\sqrt{p(s_{(b)}; \lambda)} / \partial \lambda_i \}_{b=1}^B$ by backpropagation for the Transformer
            \STATE Estimate $\{ O_{\lambda_i}(s_{(b)}) = i \partial \gamma(s_{(b)}; \lambda) / \partial \lambda_i \}_{b=1}^B$ by backpropagation for the phase neural network
            \STATEx \hspace{4em}\COMMENT{Use quantum computers for parameter shift rule}
            \STATE Estimate $\{ O_{\theta_i}(s_{(b)}) = (f[s_{(b)}; U(\theta_i + \pi/2)] - f[s_{(b)}; U(\theta_i - \pi/2)])/2 \}_{b=1}^B$ for the amplitude part circuit. For the phase part, multiply by an imaginary unit
            \STATE Estimate $\{ O_{c_i}(s_{(b)}) = \bra{s} U^\dag(\theta) Z_i U(\theta) \ket{s} \}_{b=1}^B$ on a quantum computer for the amplitude part circuit. For the phase part, multiply by an imaginary unit
            \STATEx \hspace{4em}\COMMENT{Obtain the variational energy, gradients and quantum Fisher matrix}
            \STATE Calculate the estimation of $E_{\rm VMC}^H$ using Eq.~(\ref{eq:E_VMC})
            \STATE Calculate the estimation of $F_i$ and $S_{i,j}$ for all $i(j)$ using Eqs.~(\ref{eq:F}) and (\ref{eq:S}) to obtain $F$ and $S$
        \ENDFOR
        \STATE Update all the parameters: $\mathcal{W} \gets \mathcal{W} - \eta (S + \epsilon I)^{-1} F$, or with the Adam optimizer
    \ENDFOR
    \STATE {\bf Return} $E_{\rm VMC}^H$
\end{algorithmic}
\end{algorithm}
\end{figure}

\section{Performance of the hybrid quantum-neural networks for a AFH chain}\label{app:AFH}
\begin{figure}[htbp]
\centering
\includegraphics[width=0.5\linewidth]{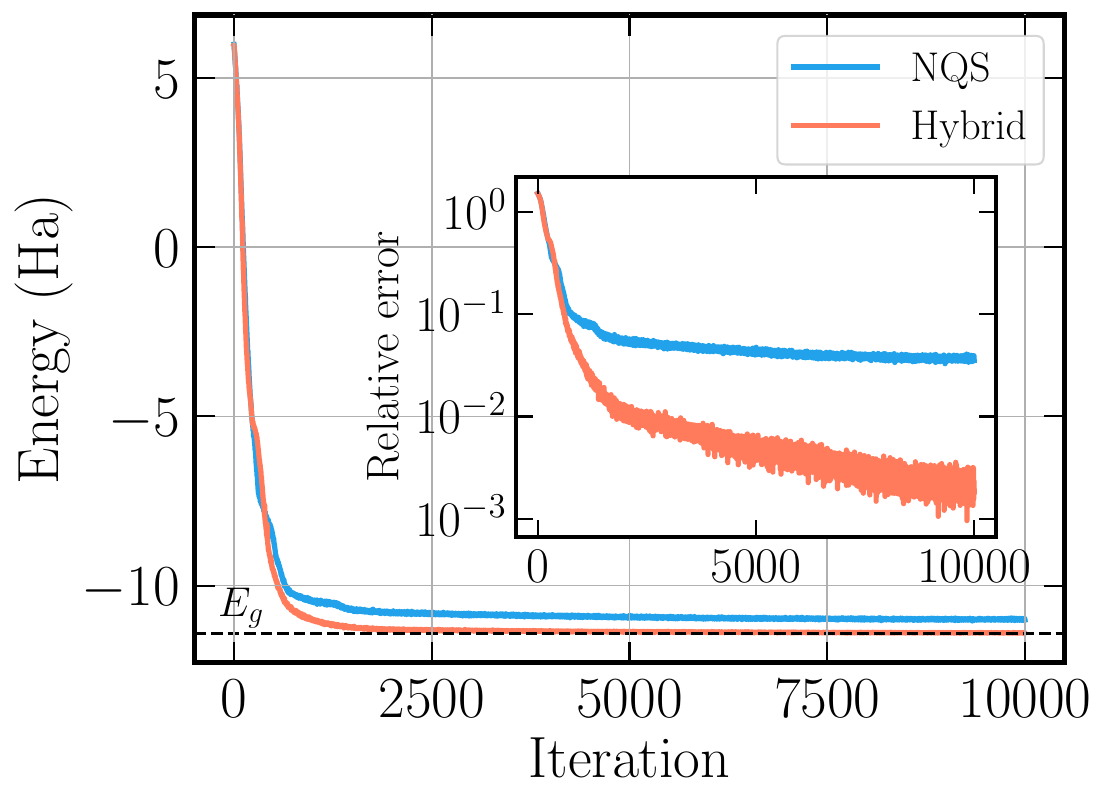}
\caption{Energy convergence curve for the NQS (blue) and the hybrid quantum-neural wavefunction (orange). For both algorithms, the sample size is fixed at $B=10^4$. In the hybrid method, the quantum circuit measurements are simulated with $M=10^4$ shots. The model being solved is a $7$-spin AFH chain with PBC. The Transformer has an embedding dimension of $d=4$, $h=2$ heads and $T=1$ block and the feedforward neural network has two hidden layers with $16$ and $8$ neurons, respectively. The quantum circuit consists of $4$ layers employing a full entanglement strategy. The Adam optimizer is used for training.}
\label{fig:hybrid_vs_nqs-Heisenberg-Nq=7}
\end{figure}

As a supplement to Fig.~\ref{fig:hybrid_vs_nqs-LiH-distance=2.4}, Fig.~\ref{fig:hybrid_vs_nqs-Heisenberg-Nq=7} shows the energy optimization curves of the NQS and our hybrid quantum-neural networks for a 7-spin AFH chain with PBC. Similar trends are observed. The Transformer comprises 290 parameters, the neural network for the phase component contains 272 parameters, and the quantum circuit includes 70 gate parameters. By the end of the optimization process, the hybrid wavefunction achieves a relative error of nearly $10^{-3}$, whereas the NQS shows a relative error exceeding $3 \times 10^{-2}$. Notably, increasing the embedding dimension of the Transformer from $4$ to $8$ yields comparable accuracy, although the number of Transformer parameters grows to $\mathcal{O}(10^3)$. 

\section{Detailed hyperparameters for calculating the ground state of LiH}\label{app:parameters}

\begin{table}[htbp]
\begin{tabular}{|c|l|l|}
\hline
\multirow{5}{*}{\shortstack{Transformer neural \\  network}} & Qubit size              & 2                 \\ \cline{2-3} 
                                            & Position size           & $N_q+1$           \\ \cline{2-3} 
                                            & Embedding dimension $d$    & 3                 \\ \cline{2-3} 
                                            & Number of heads $h$        & 1                 \\ \cline{2-3} 
                                            & Number of blocks $T$       & 1                 \\ \hline
\multirow{2}{*}{\shortstack{Feedforward \\ neural  network}}       & Number of hidden layers & 2                 \\ \cline{2-3} 
                                            & Neurons per layer       & $[16,8]$       \\ \hline
\multirow{3}{*}{Quantum circuit}            & Number of layers $N_l$       & 4                 \\ \cline{2-3} 
                                            & Entanglement strategy        & Full              \\ \cline{2-3} 
                                            & Number of shots $M$         & $10^4$            \\ \hline
\multirow{2}{*}{Adam optimizer}             & First moment decay rate $\beta_1$               & 0.9               \\ \cline{2-3} 
                                            & Second moment decay rate $\beta_2$               & 0.95              \\ \hline
\multirow{2}{*}{Cosine annealing}           & Initial learning rate $\eta_{init}$   & $5\times 10^{-3}$ \\ \cline{2-3} 
                                            & Minimal learning rate $\eta_{min}$   & $5\times 10^{-4}$ \\ \hline
\multirow{2}{*}{VMC}                        & Sample size $B$            & $10^4$            \\ \cline{2-3} 
                                            & Number of iterations $N_{iters}$                  & $3\times 10^3$    \\ \hline
\end{tabular}
\caption{Hyperparameters of the hybrid quantum-neural networks used to calculate the ground state of {\rm LiH}. In the cosine annealing schedule~\cite{loshchilov2017sgdr}, the learning rate is dynamically adjusted as $\eta_{min} + 0.5 * (\eta_{init} - \eta_{min}) * [1 + \cos{(\pi t / N_{iters})}]$, where $t$ is the current training step. The NQS does not involve quantum circuits, and all other hyperparameters are the same as those listed in this table, except that $\beta_2$ is set to $0.99$.}
\label{tab:hyperparameters}
\end{table}

\end{document}